\documentclass[aps,amsar,prb,twocolumn,superscriptaddress,showpacs,amsfonts,amssymb,amsmatht,prb]{revtex4}
\usepackage[dvips]{graphicx}
\DeclareGraphicsExtensions{.ps,.eps,.pdf}
\DeclareGraphicsExtensions{.jpg,.png}
\usepackage{color}
\usepackage{ulem}
\usepackage[colorlinks=true,linkcolor=blue,citecolor=blue,urlcolor=blue]{hyperref}
\usepackage{tabularx}
\usepackage{multirow}

\begin{document}

\title{RIXS studies of magnons and bimagnons in the lightly doped cuprate La$_{2-x}$Sr$_{x}$CuO$_{4}$}

\author{L. Chaix}
\email[Corresponding author: ]{laura.chaix@cea.fr}
\thanks{Present address: CEA, Centre de Saclay, /DSM/IRAMIS/ Laboratoire L\'eon Brillouin, 91191 Gif-sur-Yvette, France.}
\affiliation{Stanford Institute for Materials and Energy Sciences, SLAC National Accelerator Laboratory and Stanford University, 2575 Sand Hill Road, Menlo Park, California 94025, USA.}
\author{E. W. Huang}
\affiliation{Stanford Institute for Materials and Energy Sciences, SLAC National Accelerator Laboratory and Stanford University, 2575 Sand Hill Road, Menlo Park, California 94025, USA.}
\author{S. Gerber}
\affiliation{SwissFEL and Laboratory for Micro and Nanotechnology, Paul Scherrer Institut, 5232 Villigen PSI, Switzerland.}
\author{X. Lu}
\affiliation{Swiss Light Source, Paul Scherrer Institut, 5232 Villigen PSI, Switzerland.}
\author{C. Jia}
\affiliation{Stanford Institute for Materials and Energy Sciences, SLAC National Accelerator Laboratory and Stanford University, 2575 Sand Hill Road, Menlo Park, California 94025, USA.}
\author{Y. Huang}
\affiliation{Swiss Light Source, Paul Scherrer Institut, 5232 Villigen PSI, Switzerland.}
\author{D.E. McNally}
\affiliation{Swiss Light Source, Paul Scherrer Institut, 5232 Villigen PSI, Switzerland.}
\author{Y. Wang}
\affiliation{Stanford Institute for Materials and Energy Sciences, SLAC National Accelerator Laboratory and Stanford University, 2575 Sand Hill Road, Menlo Park, California 94025, USA.}
\author{F. H. Vernay}
\affiliation{Laboratoire PROMES-CNRS (UPR-8521), Universit\'e de Perpignan Via Domitia, Rambla de la thermodynamique, Tecnosud, 66100 Perpignan, France.}
\author{A. Keren}
\affiliation{Department of Physics, Technion - Israel Institute of Technology, Haifa, 32000, Israel.}
\author{M. Shi}
\affiliation{Swiss Light Source, Paul Scherrer Institut, 5232 Villigen PSI, Switzerland.}
\author{B. Moritz}
\affiliation{Stanford Institute for Materials and Energy Sciences, SLAC National Accelerator Laboratory and Stanford University, 2575 Sand Hill Road, Menlo Park, California 94025, USA.}
\author{Z.-X. Shen}
\affiliation{Stanford Institute for Materials and Energy Sciences, SLAC National Accelerator Laboratory and Stanford University, 2575 Sand Hill Road, Menlo Park, California 94025, USA.}
\affiliation{Geballe Laboratory for Advanced Materials, Stanford University, Stanford, California 94305, USA.}
\author{T. Schmitt}
\affiliation{Swiss Light Source, Paul Scherrer Institut, 5232 Villigen PSI, Switzerland.}
\author{T. P. Devereaux}
\affiliation{Stanford Institute for Materials and Energy Sciences, SLAC National Accelerator Laboratory and Stanford University, 2575 Sand Hill Road, Menlo Park, California 94025, USA.}
\affiliation{Geballe Laboratory for Advanced Materials, Stanford University, Stanford, California 94305, USA.}
\author{W.-S. Lee}
\email[Corresponding author: ]{leews@stanford.edu}
\affiliation{Stanford Institute for Materials and Energy Sciences, SLAC National Accelerator Laboratory and Stanford University, 2575 Sand Hill Road, Menlo Park, California 94025, USA.}

\date{\today}

\begin{abstract}
We investigated the doping dependence of magnetic excitations in the lightly doped cuprate La$_{2-x}$Sr$_{x}$CuO$_{4}$ via combined studies of resonant inelastic x-ray scattering (RIXS) at the Cu $L$$_{3}$-edge and theoretical calculations. With increasing doping, the magnon dispersion is found to be essentially unchanged, but the spectral width broadens and the spectral weight varies differently at different momenta. Near the Brillouin zone center, we directly observe bimagnon excitations which possess the same energy scale and doping dependence as previously observed by Raman spectroscopy. They disperse weakly in energy-momentum space, and are consistent with a bimagnon dispersion that is renormalized by the magnon-magnon interaction at the zone center. 

\end{abstract}


\maketitle

%%%%%%%%%%%%%%%%%%%%%%%%%%%%%%%%%%%%%%%%%%%%%%%%%

\section{INTRODUCTION}
The parent compound of the superconducting cuprates is a spin $1/2$ antiferromagnetically-ordered (AFM) insulator, implying that the associated spin fluctuations could play a role in the pairing mechanism of superconductivity \cite{Ofer2006, Scalapino2012, Anderson2016}. Along this line, many studies have been devoted to probe the magnetic excitations in cuprates using Raman spectroscopy \cite{Devereaux2007} and inelastic neutron scattering (INS) \cite{Bourges2000, Tranquada2004, Hayden2004, Wakimoto2007}. In the last decade, the advance of resonant inelastic x-ray scattering (RIXS) has enabled sufficient resolution to study the magnetic excitations in cuprates \cite{Ament2011, Braicovich2009, Braicovich2010, LeTacon2011, Dean2013}. While INS measurements primarily focus on the low energy excitations near the AFM wavevector (0.5, 0.5)  (in reciprocal lattice units (r.l.u.), defined by ($\frac{2\pi}{a}$, $\frac{2\pi}{b}$)), RIXS complementarily probes the high energy magnetic excitations over a wider range of the Brillouin zone away from (0.5, 0.5) and is limited only by the momentum of the soft x-ray photons. In addition, RIXS can measure small samples and thin films, allowing to explore new regimes of the phase diagram where large single crystals are unavailable for INS measurements \cite{Dean2012}.

\begin{figure}
\resizebox{8.6cm}{!}{\includegraphics{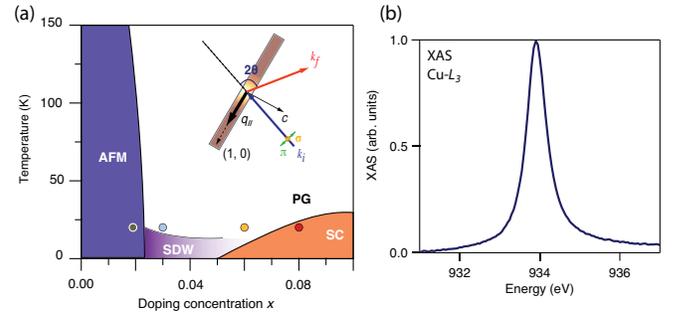}}
\caption{(Color online) 
(a) Sketch of the phase diagram in the lightly doped regime of La$_{2-x}$Sr$_{x}$CuO$_{4}$. The four sample dopings probed in our study, $x$ = 0.019, 0.03, 0.06 and 0.08, are indicated as colored round markers. The AFM, SDW, SC, and PG stand for antiferromagnetic, spin density wave, superconductivity, and pseudogap, respectively. Insert: scattering geometry of our RIXS experiment. $\sigma$ ($\pi$) represents the linear vertical (horizontal) polarisation of the incident x-ray beam. (b) X-ray absorption spectrum (XAS) of the $x$ = 0.06 sample in the $\sigma$-polarisation configuration.}
\label{fig1}
\end{figure}

Among the literature of magnetic excitations in cuprates, important issues need to be clarified. First, RIXS measurements have shown that the magnon-like magnetic excitations, the so-called paramagnon, persist well beyond the AFM phase boundary \cite{LeTacon2011, Dean2013, LeTacon2013} -- even in the heavily overdoped regime beyond the superconducting (SC) dome \cite{Meyers2017, Wakimoto2015}. Upon hole doping, the paramagnon dispersion along the $h$-direction [\textit{i.e.} (0, 0)-(1, 0)] is essentially unchanged throughout the phase diagram. In contrast, RIXS spectra along the other high symmetry direction $hh$ [\textit{i.e.} (0, 0)-(1, 1)] show a significant softening with increasing doping \cite{Meyers2017}. Such an anisotropic doping evolution appears to be inconsistent with a recent INS result and raises questions regarding the extent to which the RIXS spectrum is affected by charge excitations \cite{Wakimoto2015}. Indeed, signatures of increasing charge excitations with increasing doping have been reported, which manifest as a fluorescent-like component in the RIXS spectrum \cite{Huang2016, Minola2017}. Clarification of this open question can be provided by a comprehensive study of magnetic excitations in the lightly doped region near the SC-AFM phase boundary in which the influence of charge excitations is expected to less pronounced than in compounds of higher doping concentrations. 

Second, previous INS measurements on La$_{2-x}$Sr$_x$CuO$_{4}$ (LSCO) have revealed an incommensurate spin density wave (SDW) along the $hh$-direction near the AFM wave vector (0.5, 0.5) between the AFM and SC phases [see Fig. \ref{fig1} (a)]. Remarkably, once the system is further doped to become a superconductor, the direction of the spin incommensuration rotates by 45 degrees (\textit{i.e.} now along the $h$- or $k$-direction) and a spin gap opens that depletes spin excitations at low energy \cite{Wakimoto2000,Matsuda2002,Birgeneau2006,Fujita2012}. It is of great interest to investigate whether there is any signature in the paramagnons that can be associated with the spin incommensuration rotation near the SDW-SC phase boundary. 

The third issue regards bimagnon excitations, which were first identified in the $B_{1g}$ channel of Raman spectra \cite{Devereaux2007}. Taking LSCO as an example, it possesses an energy scale of approximately 390 meV, which is lower than that of non-interacting magnon 4$J$ due to magnon-magnon interaction ( $J$ is the superexchange interaction between nearest spins $\sim$ 120 meV in LSCO).  In addition, its spectral weight diminishes rapidly with increasing hole-doping \cite{Sugai2003,Sugai2013,Devereaux2007}. Extending to finite momentum in the reciprocal space, combined RIXS studies at both the Cu $L$-edge and the O $K$-edge reported that the dominant bimagnon branch is maximal at the zone center \cite{Bisogni2012}. However, the renormalized bimagnon dispersion due to the magnon-magnon interaction should exhibit a minimum at the zone center \cite{Vernay2007}. Notably, the bimagnon energy scale ($\sim$ 450 meV) extracted from the O $K$-edge RIXS is higher than that measured by Raman spectroscopy, casting doubt on its attribution as bimagnons. Clarification of this issue is required for a complete picture of the magnetic excitations in cuprates. 
 
To address these points, we present a Cu $L_3$-edge RIXS study to explore the evolution of magnetic excitations in lightly doped La$_{2-x}$Sr$_{x}$CuO$_{4}$ when the system changes from the antiferromagnetic to the superconducting phases. We observe that while the dispersion of the (para)magnon is insensitive to the doping, their width and spectral weight do change progressively. Near the Brillouin zone center, we identify bimagnon excitations, which disperse weakly as a function of momentum. These excitations rapidly broaden and become un-resolvable with increasing doping, consistent with the behavior of bimagnons measured by Raman spectroscopy \cite{Devereaux2007, Moritz2011}. Finally, we also compare our data with calculated magnons and bimagnons in the Hubbard model and find good agreements between the theories and experiments.

This article is structured as follows. The sample preparation, experimental method and data analysis are presented in Sec. II. The results and discussion are presented in Sec. III that is divided into two parts: Subsection A focuses on the doping dependence of the magnons, and Subsection B discusses the observation of bimagnons. Finally, Sec. IV summarizes the work.

%%%%%%%%%%%%%%%%%%%%%%%%%%%%%%%%%%%%%%%%%%%%%%%%%

\section{EXPERIMENTAL METHODS}

\begin{figure}
\resizebox{8.6cm}{!}{\includegraphics{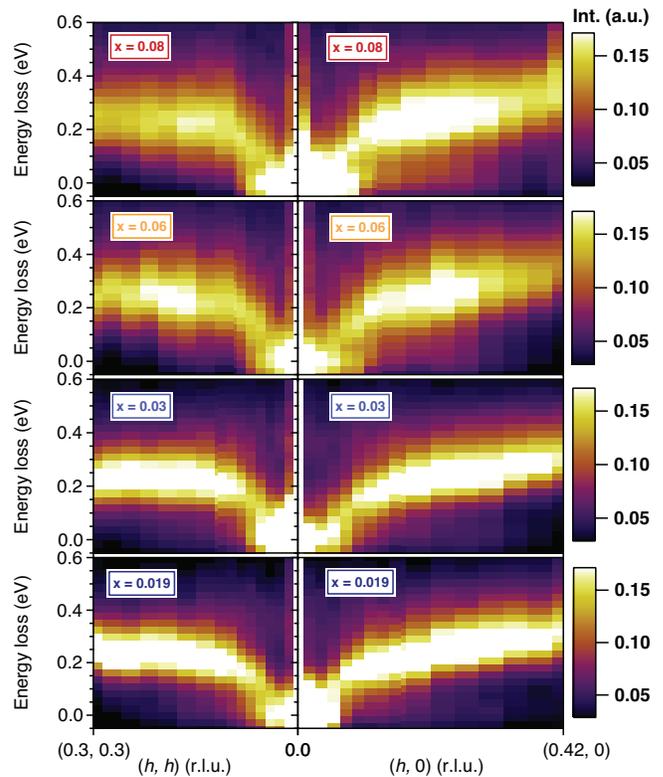}}
\caption{(Color online) RIXS intensity map along both the $h$- and $hh$-directions in the $\pi$-polarisation configuration of four heavily underdoped samples of LSCO.}
\label{fig2}
\end{figure}

\begin{figure*}
\resizebox{16.5cm}{!}{\includegraphics{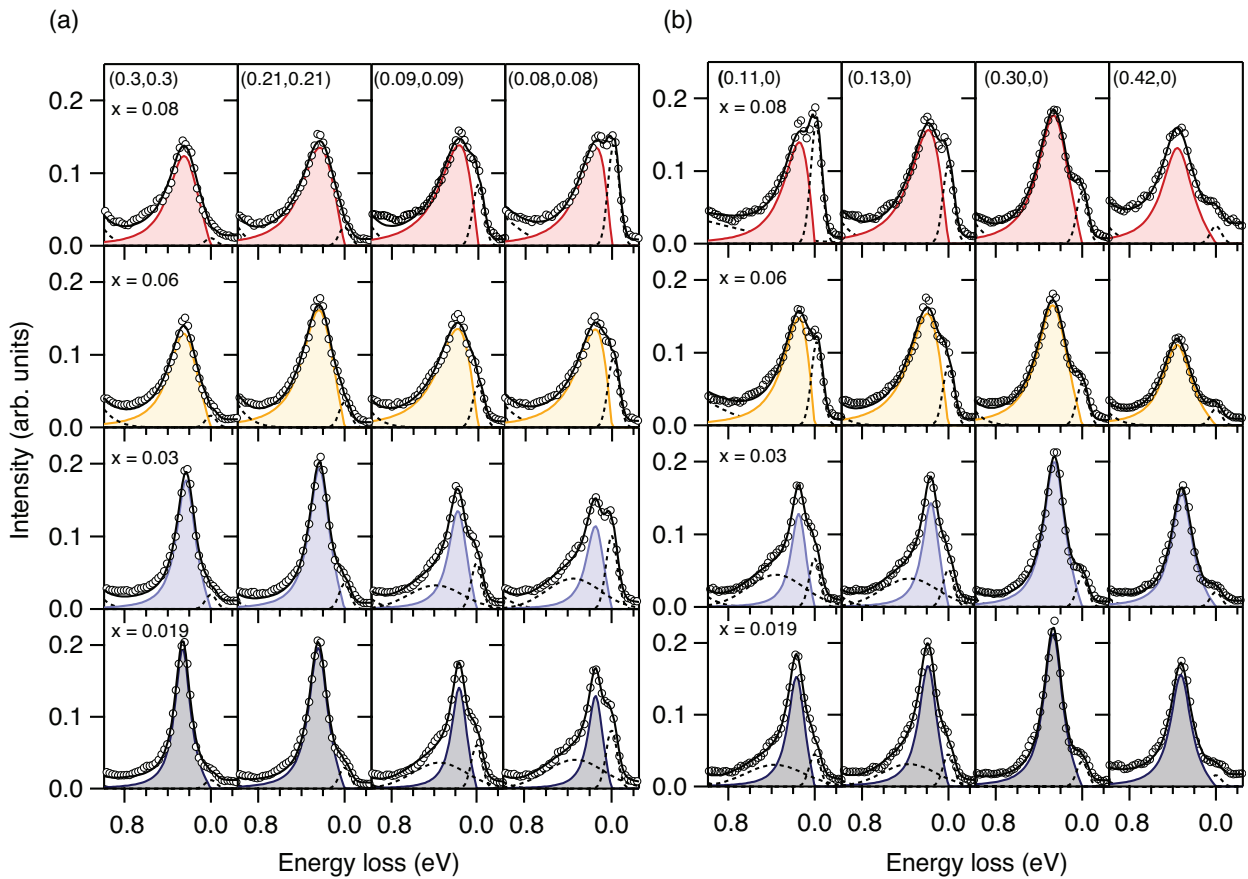}}
\caption{(Color online) Representative RIXS spectra recorded for LSCO along both the (a)$hh$- and (b)$h$-directions in the $\pi$-polarisation configuration. The corresponding fits (solid lines) are superimposed to the raw data (black circles). The fitted model consists of a fixed-width Gaussian for the elastic peak (black dashed lines) and an anti-symmetrized Lorentzian for the magnetic excitations (filled area) combined with a Gaussian background (black dashed lines) to account for the tail of the $dd$-excitations. The fitting model for the spectra near the zone center ($q_{\parallel} <$  0.1 r.l.u.) of the $x$ = 0.019 and 0.03 samples includes an additional Gaussian peak (black dashed lines) in order to account for the bimagnon excitations.}
\label{fig3}
\end{figure*}

High quality La$_{2-x}$Sr$_x$CuO$_4$ single-crystals were grown by the floating-zone method. We measured samples at four doping concentrations in the lightly doped regime of LSCO: $x$ = 0.019, 0.03, 0.06 and 0.08, as also indicated in Fig. \ref{fig1} (a). RIXS experiments were performed at the ADRESS beamline \cite{Stocov2010} of the Swiss Light Source at the Paul Scherrer Institute using the SAXES spectrometer \cite{Ghiringhelli2006}. The samples were characterized, cut and aligned using Laue diffraction prior to our RIXS measurements. To have a clean $a, b$-plane surface, the samples were cleaved inside the vacuum chamber (better than $10^{-8}$ torr) right before the RIXS measurements. 

In our study, all RIXS spectra were recorded with the incident photon energy tuned to the maximum of the absorption curve at the Cu $L_3$-edge [see Fig. \ref{fig1} (b)]. The total energy resolution was approximately 120 meV and the scattering angle was set to 2$\theta$ = 130$^\circ$, to maximize the momentum transfer. We assume that the magnetic excitations are quasi-two dimensional ($i.e.$ no dispersion along the $c$-axis); thus the dispersions are plotted as a function of the projected in-plane momentum transfer $q_{\parallel}$ [see insert of Fig. \ref{fig1} (a)]. Dispersions along the two high symmetry directions $h$- and $hh$- were measured. Data were taken at $T$ = 20 K using either linear vertical ($\sigma$) or horizontal ($\pi$) polarisations of the incident x-ray beam [see insert of Fig. \ref{fig1} (a)] depending on the nature of investigated excitations. In our convention, the spectrum of positive $q_{\parallel}$ (\textit{i.e.} grazing-emission geometry) in the $\pi$-polarisations configuration is dominated by magnon excitations \cite{Peng2015}. 

The zero-energy alignment is first coarsely determined by the elastic peak position of a carbon tape that is mounted next to the samples. The zero energy was finely adjusted during the fitting procedure of the RIXS spectra. Following previous works on magnon excitations, the fitting model consists of Gaussian functions for the elastic peak and the bimagnon in the lowest doping concentration samples (and only near the zone center), an anti-symmetrized Lorentzian for the magnetic excitation (magnon) and a background that fits the tail of the $dd$-excitations at higher energy. We note that anti-symmetrized Lorentizan function were used to ensure that the imaginary part of the spin susceptibility is an odd function, as described in the supplementary information of a previous work \cite{LeTacon2011}. All the RIXS spectra were normalized by the spectral weight of the $dd$-excitations, as did previous RIXS studies of magnetic excitations in cuprates \cite{Braicovich2009,Braicovich2010,Dean2012,Dean2013,LeTacon2011}.

\begin{figure*}
\resizebox{16.5 cm}{!}{\includegraphics{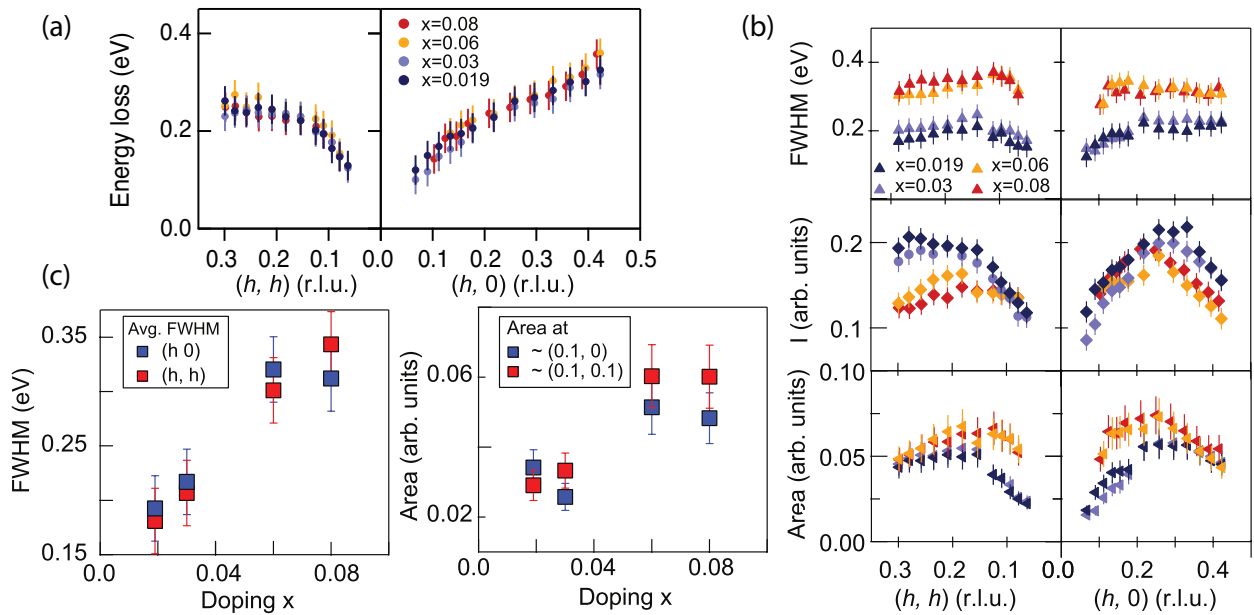}}
\caption{(Color online) (a) Doping evolution of the dispersion extracted from our RIXS data. The error bars of the dispersions are estimated by the uncertainty in determining the zero energy loss. (b) Doping evolution of the FWHM, the peak intensity $I$, and  the area of the magnetic excitations. Error bars are estimated using the uncertainty in determining the zero energy loss for the FWHM and the noise level in the data for the intensity and the area.  (c) Averaged FWHM for all momentum positions (left) and the area near the zone center (right) as a function of doping. The directions and positions in the reciprocal space that was used for FWHM average and area are indicated in the figure legends, respectively. For the ease of comparing to raw data, the FWHM and area plotted here did not deconvolve the instrument resolution from the data.}
\label{fig4}
\end{figure*}

\begin{figure}
\resizebox{8.6 cm}{!}{\includegraphics{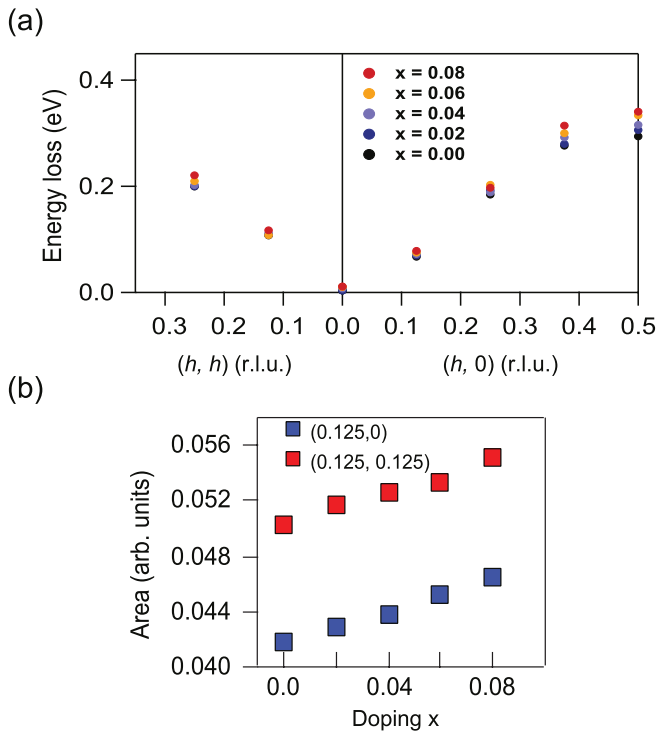}}
\caption{(Color online) (a) Peak positions of $S(\bm{q},\omega)$ from a DQMC calculation of the three-band Hubbard model on a $8 \times 8$ lattice. (b) The area, \textit{i.e.} the spectral weight of the calculated $S(q,\omega)$ by DQMC, at representative momentum positions near the zone center, is plotted as a function of the doping concentration. }

\label{fig5}
\end{figure}

\begin{figure*}
\resizebox{16.5 cm}{!}{\includegraphics{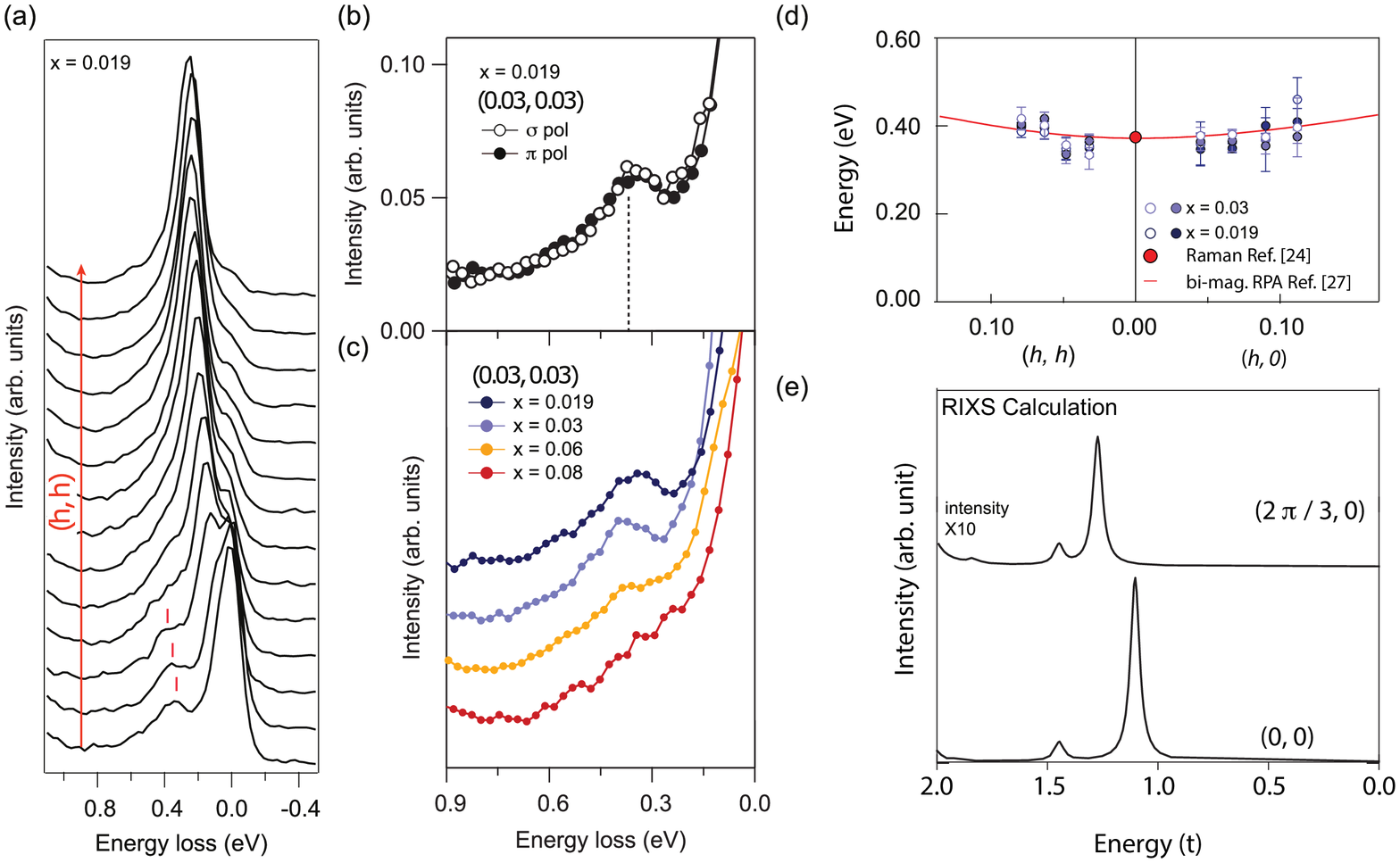}}
\caption{(Color online) (a) Stack plot of representative RIXS spectra along the $hh$-direction near the zone center, taken with the $\pi$-polarisation configuration. Bimagnon excitations are indicated by red ticks. (b) Bimagnon excitation recorded with both $\sigma$ or $\pi$ polarisations. (c) Doping evolution of the bimagnon excitation at $q_{\parallel}$ $\approx$ (0.03, 0.03), taken with the $\pi$-polarisation configuration. The spectra are offset along the intensity axis for clarity. (d) Dispersions of the bimagnon for the $x$ = 0.019 and $x$ = 0.03 samples along the $h$- and $hh$-directions. For the bimagnon analysis, the peak positions have been extracted from a different fitting procedure than the one used in Fig. \ref{fig3}, $i.e.$ for the magnon excitations: a Gaussian peak for the elastic peak, a constant background, and two anti-symmetrized Lorentzians for bimagnons and magnons, respectively. The error bars correspond to one standard deviation from the fit. The calculated bimagnon dispersion from Ref. \cite{Vernay2007} (red curve) is also superimposed by normalizing the zone center energy to the bimagnon energy measured via Raman \cite{Sugai2013}, which is also indicated as the red solid circle. (e) Exact diagonalization calculations of the RIXS spectrum. The calculation was performed on a single band Hubbard model at half-filling, \textit{i.e.} for the undoped case.}
\label{fig6}
\end{figure*}

\section{RESULTS AND DISCUSSION}

\subsection{Doping evolution of magnon excitations in the underdoped regime of LSCO}

Figure \ref{fig2} shows RIXS intensity maps of the doping concentrations we have investigated. Magnetic excitations in all samples exhibit similar energy-momentum dispersions, which reach maximal energies of approximately 0.325 eV and 0.250 eV at AFM zone boundaries (0.5, 0) and (0.25, 0.25), respectively. Raw spectra at representative momentum positions are shown in Fig. \ref{fig3}. For the $x$  = 0.019 and 0.03 samples, which are near the AFM-SDW phase boundary in the phase diagram, the spectra are dominated by one sharp peak. This peak, consistent with previous measurements \cite{Braicovich2009}, is attributed to the magnon excitation that is seen as the dispersive feature  in the RIXS intensity map (Fig. \ref{fig2}). When the doping concentration increases to just beyond the onset doping of SC, \textit{i.e.} for the $x$ = 0.06 and 0.08 data as shown in Fig. \ref{fig3}, the magnon-like peak still remains but broadens, forming the so-called ``paramagnon" \cite{LeTacon2011}.

To quantify these changes, the data were fitted to extract the peak positions, widths and intensities. As depicted in Fig. \ref{fig3}, the fits agree well with the data. The dispersions of the magnon and paramagnon excitations (\textit{i.e.} the fitted peak positions versus in-plane momentum) are shown in Fig. \ref{fig4}. We find that the dispersions of different doping concentrations along both the $h$- and $hh$-directions are essentially identical within our experimental accuracy. While previous RIXS measurements reported that the paramagnon dispersion along the $hh$- direction significantly softens in the optimally- and over-doped regime \cite{Huang2016, Meyers2017}, our results indicate that such softening does not occur in the underdoped regime near the AFM-SC phase boundaries.

The fitted full width at half maximum (FWHM) is summarized in the upper panels of Fig. \ref{fig4} (b).  In all measured samples, the widths of the magnon and paramagnon are essentially momentum independent with slightly smaller values near the zone center. As a function of the doping concentration, as shown in the left panel of Fig. \ref{fig4} (c), the averaged FWHM over all the momentum points progressively increases when increasing hole-doping. These findings are fully consistent with previous RIXS measurements over a larger doping range \cite{Dean2012,Meyers2017}. Interestingly, the spectral intensity ($i.e.$ fitted peak height) exhibits a momentum-dependent variation as a function of doping. As shown in the middle panels of Fig. \ref{fig4} (b), while the intensity is essentially doping independent at small momenta near the zone center, the intensity reduces with increasing doping at larger momenta near the zone boundaries. As a consequence, at large momentum transfer near the zone boundaries, the area of the magnetic spectrum, \textit{i.e.} the spectral weight, appears to be conserved as a function of doping. On the other hand, the spectral weight near the zone center increases with hole doping, as shown in the lowest panels of Fig. \ref{fig4} (b) and the right panel of Fig. \ref{fig4} (c). 

It is informative to compare our results with the spin susceptibility of the three-band Hubbard model, which realistically reflects the electronic structure of cuprates \cite{Scalettar1991,Dopf1992,Kung2016}. We use determinant quantum Monte Carlo (DQMC) to calculate the dynamical spin structure factor $S(\textit{q},\omega)$. The parameters in units of eV are $U_{dd}=8.5, U_{pp}=4.1, \Delta_{pd}=3.24, t_{pd}=1.13, t_{pp}=0.49$ and the chemical potential is used to adjust the hole doping concentration. The model is simulated on an $8 \times 8$ unit cell cluster with periodic boundaries at a temperature of $T = 0.125$ eV (\textit{i.e.} $\sim$ 1500 K). Data for the spin susceptibility are collected into 400 bins with 50000 Monte Carlo samples each. To analytically continue the spin susceptibility to $S(\bm{q},\omega)$, we use the maximum entropy method with a model function derived from the first moments of each spectra as described in \cite{Jarrell1996,Macridin2004,EHuang2016}. As shown in Fig. \ref{fig5} (a), the calculated dispersions  do not show significant changes in the dispersion of the magnetic excitations along both $h$- and $hh$-directions when the doping concentration increases from $x$ = 0 to $x$ = 0.08, consistent with our experimental results shown in Fig. \ref{fig4} (a). 

Figure \ref{fig5} (b) shows the area, \textit{i.e.} the spectral weight, of the $S(\bm{q},\omega)$ extracted from our DQMC calculations at smaller momentum transfer near the zone center. A small increase of spectral weight is also seen, but this increase is very subtle and much smaller than that seen in the RIXS data. We remark that DQMC calculations were inevitably performed at high temperatures due to calculation efficiency, which may underestimate the spectral weight changes at low temperatures in the experiments. In addition, we note that the increase of spectral weight near the zone center in the data could be partially due to the inclusion of remnant bimagnon excitations or other charge excitations \cite{Minola2017}. Investigations using polarisation-resolved RIXS and and comparison with appropriately calculated charge contribitions \cite{Jia2013, Jia2016, Minola2017} would be necessary to provide further insight into this momentum-dependent variation.

\subsection{Bimagnon observation at the Cu $L$$_{3}$-edge}

A new discovery in our data is the direct observation of bimagnon excitations using Cu $L_3$-edge RIXS. Figure \ref{fig6} (a) shows data taken on the $x$ = 0.019 sample. An additional peak in the tail of the elastic signal is clearly visible near the zone center. With increasing momentum along both the $h$- and $hh$-directions, it weakly disperses and eventually becomes unresolvable beyond momenta larger than approximately 0.1 r.l.u., where the magnon excitations completely dominate the spectra (see also Fig. \ref{fig3}). The mode energy is found to be $E \sim$ 0.38 eV, and can be resolved using either the $\pi$- or $\sigma$-polarisation of the incident x-rays [see  Fig. \ref{fig6} (b)]. Importantly, as shown in Fig. \ref{fig6} (c), the mode rapidly diminishes with increasing doping concentration. At $x$ = 0.08, the mode is unresolvable in our data. We remark that the energy of this mode and its doping dependence are essentially identical to the bimagnon excitations observed via Raman spectroscopy \cite{Sugai2003,Sugai2013,Devereaux2007,Moritz2011}. Thus, we attribute this mode near the zone center to bimagnon excitations.

The bimagnon dispersion along both the $h$- and $hh$-directions near the zone center can be extracted for the $x$ = 0.019 and $x$ = 0.03 samples, as shown in Fig. \ref{fig6} (d). The bimagnon energy appears to increase slightly with increasing momentum transfer before the mode becomes irresolvable.  In Fig. \ref{fig6} (d), we superimpose a calculated bimagnon dispersion obtained via the random phase approximation (RPA) \cite{Vernay2007}, by normalizing its energy scale to match the bimagnon energy measured by Raman spectroscopy at the zone center [red line and marker in Fig. \ref{fig6} (d)]. In the calculation, a magnon-magnon interaction that reduces the non-interacting bimagnon energy from 4$J$ to 2.78$J$ is included. The calculated dispersion is found to be consistent with our data. Thus, our results lend support to the existence of magnon-magnon interaction and the association of the observed peaks with bi-magnon excitations from doped antiferromagnets.. 

The attribution of this excitation to bimagnon appears to contradict with an earlier theoretical work that predicts a negligible bimagnon spectral weight near the zone center in the Cu $L_3$-edge RIXS spectrum \cite{Bisogni2012}. To investigate this cross-section issue, we performed exact diagonalization calculations, using the single-band Hubbard model on a half-filled 12-site cluster with $U = 8t$, $t' = 0.3t$. This cluster mimics the parent compound of cuprates in which the bimagnon excitations are most robust and free of the complications from the charge excitations due to doped holes. RIXS $L_3$-edge spectra were calculated using the Kramers-Heisenberg formula \cite{Ament2011} with the core-hole potential $U_c =  4t$, spin-orbit coupling in the $2p$ shell $\lambda = 32.5t$, and the inverse of core-hole lifetime  $\Gamma = t$, same as those used in Ref. \cite{Jia2016}.As shown in Fig. \ref{fig6} (e), the calculated RIXS cross-section shows strongest intensities at the zone center and significantly weaker intensities at large momentum, consistent with our experimental observations. We note that the RIXS cross-section shown in previous work by Bisogni \textit{et al.} \cite{Bisogni2012} was computed under several levels of approximations: (1) an ultrashort core-hole lifetime expansion is used to simplify the Kramers-Heisenberg formula to two particle correlators, (2) a spin-only Heisenberg model instead of a Hubbard model is used and (3) linear spin wave theory is employed, where the magnon-magnon interaction has not been included. Our exact diagonalization calculations using the Hubbard model enables us to evaluate the exact RIXS cross-section non-perturbatively. We suspect that the $\sim$ 450 meV excitation observed by previous RIXS measurements at the O $K$-edge may have a different character than bimagnon excitations. \cite{Bisogni2012}.

We remark that the bimagnon excitation discussed here has a net spin change of zero ($\Delta S_z = 0$ which involves spin flips of two neighboring sites with opposite direction. The readers might wonder whether bimagnon excitation with $\Delta S_z = 2$ can be detected by the Cu $L$-edge RIXS. We note that magnetic excitations both measured and theorized for Cu $L_3$-edge (direct) RIXS have been discussed in the literature (see for example, Section V. E. in the Ref. \cite{Ament2011}).  In essence, the intermediate state of the L-edge RIXS process, in particular the spin-orbit coupling in the core, plays the key role.  Since cupartes are spin $1/2$ systems, this spin-orbit term can generate at most a single spin flip $\Delta S_z = 1$, not $\Delta S_z = 2 $. We note that this statement is specifically for the Cu $L$-edge RIXS on Cu$^{2+}$ system.  For other compounds, such as Ni$^{2+}$ spin 1 system, $\Delta S_z = 2$ excitation is possible (for example, see Ref. \cite{deGroot1998}, Ref. \cite{Ghiringhelli2009}, and also Section V. E. in the Ref. \cite{Ament2011}). We remark that higher-order terms in the scattering cross-section beyond the 2nd order Kramers-Heisenberg formula for resonant scattering (itself an approximation to Fermi's Golden rule) may possess processes with $\Delta S_z = 2$ in Cu $L$-edge RIXS, but with little contribution to the overall spectral intensity. 

\section{Summary}

To clarify the three issues listed in the introduction, we have studied the magnons and bimagnons in the heavily underdoped regime of La$_{2-x}$Sr$_{x}$CuO$_{4}$ using RIXS at the Cu $L$$_{3}$-edge in the energy-momentum space away from the AFM wave-vector (0.5, 0.5). First, we have shown that the dispersion of the magnons does not change with doping neither along the $h$- nor the $hh$ directions. The width exhibits a progressive broadening with increasing doping, accompanied with a momentum-dependent variation of the intensity and spectral weight. These observations are consistent with a recent neutron scattering study \cite{Wakimoto2015} that the magnetic excitation does not exhibit strong softening even up to overdoped regime. This is also consistent with calculations from Hubbard model.

Second, concerning the spin incommensuration near the AFM wavevector (0.5, 0.5), which is known to rotate by 45$^{\circ}$ when the system is doped across the SDW-SC phase boundary, we do not resolve a corresponding sudden change in paramagnon in the similar doping range. However, since our measurement temperature (20 K) is comparable to the onset temperature of SDW ( 20  $\sim$ 30 K), lower temperature might be needed to resolve the signature in paramagnon that is associated with the spin-incommensuration rotation at (0.5, 0.5). Nevertheless, our results support that the magnetic excitations near (0.5, 0.5) indeed exhibit the most dramatic variation in response to doping, and thus, are most relevant to the underlying quantum phases emergence in cuprates \cite{Huang2017}.

Finally, concerning the third issue, we observed bimagnon excitations near the zone center in the energy-momentum space that possess an energy scale and doping dependence consistent with those seen via Raman spectroscopy. Our calculation indicates that bimagnon excitations do possess non-zero cross-section near the zone center in Cu $L_3$-edge RIXS, further supporting our experimental observation of bimagnons. The dispersion is found to be consistent with the renormalized bimagnon branch due to magnon-magnon interaction, as proposed in previous work. \cite{Vernay2007}

Our results complement previous measurements \cite{LeTacon2013,Dean2013,Meyers2017} by providing missing information in the lightly doped regime of the phase diagram, allowing a more complete picture of magnetic excitations in cuprates. We remark that our results allow for quantitative assessment of the calculations of the Hubbard for the spin response that can be directly compared to the data. The current agreement indicates that the spin excitations and their doping dependence are quite adequately reproduced by simulations. Notably, the correlations of stripe phase have been recently found in the doped Hubbard model using state-of-the-art numerical computation \cite{Huang2017Science, Zheng2017}, and a modification of paramagnon at the charge order wave vector has also been reported in a stripe-ordered cuprate \cite{Miao2017}. It is then an intriguing question of whether a similar agreement could be found when extending these methods into the stripe and superconducting state. This could provide more detailed information on whether the ground state of the doped Hubbard model truly is superconducting.

%%%%%%% References

\section{Acknowledgments}
The experimental work was performed at the ADRESS beamline of the Swiss Light Source at the Paul Scherrer Institut. This research was supported by the Division of Materials Sciences and Engineering of the U.S. Department of Energy, Office of Basic Energy Sciences, under contract DE-AC02-76SF00515 at the SLAC National Accelerator Laboratory. T.S. acknowledged supports by the Swiss National Science Foundation through its Sinergia network Mott Physics Beyond the Heisenberg Model (MPBH) and the NCCR MARVEL. X. L. acknowledges financial support from the European Community's Seventh Framework Program (FP7/2007-2013) under Grant Agreement No. 290605 (COFUND: PSI-FELLOW).

%\bibliography{LSCO_biblio}

\end{document}